\documentclass[sigconf]{acmart}
\setcopyright{none}
\renewcommand\footnotetextcopyrightpermission[1]{}
\usepackage{listings}
\usepackage{xcolor}

\usepackage{multirow}
\usepackage{tabularx}
\usepackage{booktabs}
\usepackage{array}
\usepackage{pifont}
\usepackage{listings}

\newcommand{\cmark}{\ding{51}} % check mark
\newcommand{\xmark}{\ding{55}} % cross mark

\usepackage[ruled,vlined]{algorithm2e}
\usepackage{tikz}

\definecolor{vgreen}{RGB}{104,180,104}
\definecolor{vblue}{RGB}{49,49,255}
\definecolor{vorange}{RGB}{255,143,102}

\lstdefinelanguage{json}{
    basicstyle=\ttfamily\footnotesize,
    breaklines=true,
    frame=single,
    showstringspaces=false
}

\lstdefinestyle{verilog-style}
{
    language=Verilog,
    backgroundcolor=\color{gray!10},
    morekeywords={assert},
    keywordstyle=\color{blue},
    identifierstyle=\color{black},
    commentstyle=\color{vgreen},
    numberstyle=\tiny\color{black},
    numbers=left, numbersep=2pt,
	xleftmargin=0.35cm, framexleftmargin=0.35cm,
    moredelim=*[s][\colorIndex]{[}{]},
    literate=*{:}{:}1
}

\lstdefinestyle{inline-verilog-style}
{
    language=Verilog,
    backgroundcolor=\color{gray!10},
    morekeywords={assert},
    keywordstyle=\color{blue},
    identifierstyle=\color{black},
    commentstyle=\color{vgreen},
    numberstyle=\tiny\color{black},
    moredelim=*[s][\colorIndex]{[}{]},
    literate=*{:}{:}1
}

\begin{document}

\title{CHARGE: Leveraging CWE Hierarchies for Hardware Security SystemVerilog Assertion Generation}


%\author{Anonymized for submission}
\author{Xiao Tan}
\affiliation{
  \institution{UNC Chapel Hill}
  \city{Chapel Hill, North Carolina}
  \country{USA}}
\email{tanxiao@unc.edu}

\author{Cynthia Sturton}
\affiliation{
  \institution{UNC Chapel Hill}
  \city{Chapel Hill, North Carolina}
  \country{USA}}
\email{csturton@cs.unc.edu}

\renewcommand{\shortauthors}{Tan et al.}
%\renewcommand{\shortauthors}{Anon.}

%% ICCAD: 8 pages, plus 1 page refs. 9-10pt. 4/7; 4/14.
\begin{abstract}

  This paper presents CHARGE, an automated framework for generating security
  properties for unverified RTL modules using CWEs and large language models
  (LLMs). The hallmark is a
  reasoning process that leverages the hierarchical nature of CWE entries to
  improve accuracy when identifying security-critical assets in unverified RTL
  modules. As a result, the approach can infer expected
  security behaviors and generate properties from identified assets and
  CWE semantics, avoiding the 
  need for trusted design specifications and reducing manual engineering
  effort. 
  We evaluate the framework on
  the Hack@DAC18, 19, and 21 open source SoC designs using OpenAI’s
  GPT-4.1. CHARGE detects 27 of 42 known bugs in these designs. For Hack@DAC21 OpenPiton SoC,
  89\% of the generated SVAs run in Cadence JasperGold FPV, and 92.2\% are
  non-vacuous. We compare to an open-source, manually written set of properties
  for these designs and find that CHARGE correctly writes properties for three
  bugs in which the manually written properties were incorrect. In addition,
  CHARGE-generated properties identify a new bug in the Hack@DAC21 OpenPiton SoC that was not previously identified.

\end{abstract}




%% \received{20 February 2007}
%% \received[revised]{12 March 2009}
%% \received[accepted]{5 June 2009}

\maketitle

% Add for arXiv submission
\begin{center}
\large\bfseries
This paper is an extended version of the paper accepted to\\
the IEEE/ACM International Conference on Computer-Aided Design (ICCAD 2026).
\end{center}

\vspace{1em}
% Add for arXiv submission
\pagestyle{plain}

\section{Introduction}

Insecure hardware can lead to compromise of the entire
system~\cite{10.5555/3361338.3361354}. To protect against this outcome, security
verification early in the design life-cycle aims to find
weaknesses in the design that could be exploitable post-deployment.
Verification techniques differ, but many rely on finding assertion failures in
the design. For example, model checking can prove a design will never
violate an assertion~\cite{symbiyosis}. While less formal techniques, such as symbolic
execution~\cite{ryan2023sylvia,Ryan2025Sylq} or fuzzing~\cite{Laeufer2018Mux,miftah2025symbfuzz,Muduli2020Hyperfuzzing} can
find instances of assertion violations in large designs.

Unfortunately, the verification outcomes are only as strong as the set of assertions used,
and writing a comprehensive set of assertions is a challenging and
time-consuming task~\cite{10.1145/3489517.3530637}. In the past, techniques
involving behavior mining from traces of
execution~\cite{goldmine2010}, translating properties from one design to another~\cite{zhang2020transys}, or using a set
of templates to generate properties~\cite{restuccia2021aker} have been used. More recently, a body of
work is emerging exploring the use of large language models (LLMs) to generate
assertions~\cite{kande2024security}.

Two major challenges to using LLMs to generate assertions persist. The first is
that LLMs need quite detailed guidance in order to produce meaningful and
accurate assertions~\cite{kande2024security}. The second is that if the goal is to use the
assertions to find bugs in the design, then the design itself cannot be assumed
to be bug-free and cannot act as a clean specification to feed to the LLM. In
the context of functional verification, these challenges are being successfully
addressed by providing the LLM with the design's specification, which details
signal names, timing, and expected behavior. In the context of security
verification, however, no such design-specific specifications exist. There are
established high-level security goals, for example the Common Weakness
Enumeration (CWE) database~\cite{cwe}, but these are not design specific, and
mapping the high-level goals to the signals, timing, and behavior of a
particular design is challenging.

To address these challenges, we present CHARGE (CWE Hierarchy 
AsseRtion Generation Engine), an automated framework that leverages the 
hierarchical structure of the CWE database to guide LLMs to generate security assertions 
directly from unverified RTL. We use the CWE entries in lieu of a specification
and use a three-step framework to guide the LLM. In the first step, the
hierarchy of the CWE is used to guide the LLM toward identifying the relevant
assets (signals) in the given RTL. In the second step, the target CWE is used to guide
the LLM toward naming the expected behavior of the assets. In the third step,
the gathered information is used to prompt the LLM to generate a SystemVerilog
Assertion (SVA) that would catch a violation of the target CWE.

We compare the assertions generated by our framework 
with those manually written for recent Hack@DAC SoC designs (2018~\cite{pulpissimo2018hack}, 2019~\cite{cva62019hack}, and 2021~\cite{openpiton2021hack}) and provided in the open-source
Verification Benchmark repository~\cite{rogers2025hardware}. 
In our experiments CHARGE generates assertions that detect 
27 out of 42 bugs. For assertions generated on Hack@DAC 2021 SoC, 89\% of the 
assertions can be directly used in the JasperGold FPV environment without 
syntax errors or using invalid signals. 
For the 15 undetected bugs, we analyze the failure cases and find that some
target CWEs tend to be more difficult for the LLM, which aligns with prior
findings~\cite{ahmad2025lashedllmsstatichardware} and that longer RTL modules are more challenging as
well. Moreover, CHARGE generates an assertion that finds an undocumented bug in Hack@DAC 2021's 
OpenPiton SoC, and we reported this issue to Hack@DAC 2021's organizers. We also 
find three instances where CHARGE produces a correct property, but the manually
written property in the Verification Benchmark repository was incorrect, 
and we reported this issue to the maintainers of the repository. 
%%%%%%%%% Discussion
Lastly, we discuss the position of CHARGE in the context of related work, technical considerations, and the scope and limitations of our evaluation.

Our main contributions are the following.
\begin{itemize}
\item We present a methodology to leverage the hierarchical nature of CWEs to
  address the challenges in generating security SystemVerilog Assertions
  directly from unverified RTL.

\item We implement the methodology in CHARGE, an automated framework that takes as input a
  target CWE and an unverified RTL module, and generates a set of SVAs.

\item We evaluate our framework on the Hack@DAC 2018, 2019, and 2021 SoCs, and
  compare to open-source, manually written properties. CHARGE generates assertions that
  detects 27 out of 42 bugs, detect one previously unreported bug, and corrects
  three of the manually written properties.
\end{itemize}

\section{Background}

\subsection{Assertioned Based Verification}
Assertion-based verification (ABV) is a widely adopted technique in hardware 
verification, which uses assertions to specify the expected behavior and 
check the correctness of the design under verification. 
SystemVerilog Assertions (SVA) is a commonly used property specification 
language in hardware verification, which allows designers to specify properties 
that can be expressed in terms of linear temporal logic. SVA can be used to 
specify both functional and security properties of a design. Functional 
properties usually originated from the design specification, while security 
properties are often highly dependent on the professional expertise of the 
engineer, as there is no formal specification for security properties of a 
design. Therefore, generating security SVA is often more challenging than 
generating functional SVA. SVAs are often manually written by engineers, which 
is time-consuming. Therefore, there is a growing interest in automatically 
generating SVA, both functional and security, to improve the efficiency of 
hardware verification. 

\subsection{LLMs for SVA generation}
LLMs show great potential in understanding natural 
language, and have been applied in code generation and analysis. Recently, 
there are research efforts in using LLMs to automatically generate SVA, 
which can significantly reduce the manual engineering effort. However, there 
are non-trivial challenges existing in this process. First, the generated 
SVA needs to be syntactically correct and able to be directly used in 
verification tools. This is challenging becasue the commercial LLMs are not 
heavily trained on Hardware description language (HDL) and SVA, and a lot 
of engineering effort is required to refine the generated SVAs. Some works 
have addressed this challenge by fine-tuning LLMs on open source hardware designs, 
while this is also limited by the availability of such datasets, comparing to 
the software code and natural language dataset. Second, 
the generated SVA needs to be semantically meaningful and able to correctly 
express the expected behavior of the design. For functional SVA generation, 
there are existing works that use LLMs to generate properties from design 
specification, which serves as a ground truth of how the design should 
behave. However, for security SVA generation, there is no formal 
specification and therefore the potentially buggy RTL is the only source 
of context, which can be misleading for LLMs and embed the bug inside the 
generated SVA. Therefore, it is important to find a better way to guide 
LLMs to generate security SVA directly from the buggy RTL, which is the 
motivation of our work.

\section{Using the Common Weakness Enumeration}

The Common Weakness Enumeration database (CWE) is a standardized taxonomy and database 
of security weaknesses maintained by MITRE~\cite{cwe}. The database is widely used to describe, 
categorize, and analyze weaknesses in both software and hardware. Each CWE 
entry includes a unique ID, a name, a description of the 
weakness, possible mitigations, and examples of vulnerable implementations. CWE
entries do not describe particular vulnerabilities in a particular
codebase. Rather, they describe a type of weakness to be avoided.

CHARGE uses the CWE database as a specification of desirable security properties. 
However, the name and description of each CWE entry are in natural language, 
which cannot be directly used for verification. Therefore, we need to 
decompose the CWE descriptions into a structured format that can be used for 
property generation. In this section, we will explain how we decompose the 
CWE descriptions into a 3-tuple format, how we select the CWEs covered by our 
framework, and how we build the CWE tree based on the hierarchy. 

\subsection{Extracting Structured Data from a CWE}

Each CWE entry describes a weak or vulnerable pattern in code. By extracting the
data from each CWE entry we can specify an associated property. We use a 3-tuple
to represent the specified security property: $\langle
\text{target},\ \text{action},\ \text{condition} \rangle$, where
\begin{itemize}
\item target is the sensitive hardware asset being protected, such as cryptographic key, data, memory region;
\item action is the operation performed on the target, such as read, write,
  zeroize; and
\item condition is the set of constraints on which the action is prohibited or allowed, typically composed of signals such as privilege level, lock bits, and status flags.
\end{itemize}

For example, CWE-226: Sensitive Information in Resource Not Removed Before Reuse
has the following description: ``The product releases a resource such as memory
or a file so that it can be made available for reuse, but it does not clear or
zeroize the information contained in the resource before the product performs a
critical state transition or makes the resource available for reuse by other
entities.'' 
The structured 3-tuple for this CWE is: $$\langle \text{sensitive resources}, \text{sanitization}, \text{reuse/release condition}
\rangle$$ And, a template SVA for this CWE is:
\begin{lstlisting}[
    caption={Template SVA for CWE-226},
    label={lst:sva},
    float={tbph},
    style={verilog-style}]
property CWE_226;
    @(posedge clk)
    (reset_trigger_event) |-> or |=>
      (resource_sensitive_data == '0)
endproperty
\end{lstlisting}   

The 3-tuple provides a structured representation of the security property encoded in the CWE description, 
and is used by CHARGE for categorizing assets and specifying the SVA templates for assertion generation. 

\subsection{CWEs in Scope}
Our goal is to produce SVAs that can be verified as part of a standard verification
process. Consequently, our framework focuses on trace 
properties that can be expressed using temporal logic. CWEs that require reasoning 
about time or power side channels, physical environmental conditions, information-flow 
properties, or software or ISA-level behavior fall outside the scope of this framework 
and are therefore excluded. 

All CWEs covered by our framework come from the CWE-1194 Hardware Design view. 
This view contains 13 categories, based on different perspectives of hardware design. 
We include only CWEs from this view 
that are designated as ``Mappable'' in the CWE database%, and treat them as leaf nodes in the CWE hierarchy
. At the time of 
writing, MITRE lists 110 hardware-related CWE weaknesses. 

We classify these CWEs into two sets. A CWE is classified as \textit{convertible} when 
its security policy can be fully represented by a single 3-tuple and mapped to an assertion 
template. A CWE is classified as \textit{partially convertible} when only part of its security 
intent can be expressed in this form and additional behavioral reasoning is required.
For example, CWE-226 is \textit{convertible} because its security intent naturally maps to a 3-tuple: 
the target is the sensitive resource, the action is sanitization, and the condition is resource release 
or reuse. In contrast, CWE-1245 (Improper Finite State Machine) is only \textit{partially convertible}. 
While valid state transitions can be represented using a 3-tuple, properties such as unreachable 
states or privilege-escalation prevention require additional reasoning beyond a single relation.
This classification only describes representability within our abstraction and does 
not change the CHARGE workflow. Both categories are processed using the same pipeline. For 
\textit{partially convertible} CWEs, the 3-tuple serves as an initial abstraction, and we leverage the 
creativity of LLMs to infer additional expected behaviors beyond those explicitly encoded in the tuple.

It is important 
to note that the inclusion of a CWE in our framework does not guarantee that a given CWE 
can be successfully verified using SVA on a specific RTL module. Whether a CWE can be verified 
is dependent on the specific RTL module and whether the weakness manifests at the RTL level 
or arises from higher-level system configurations. In Section 4.2, we discuss how we determine 
if a CWE is verifiable for a given RTL module, which is based on whether the LLM can identify 
enough asset details. 

Other CWE-based hardware verification frameworks adopt different methodologies when 
selecting the CWEs they cover. For example, Don’t CWEAT It ~\cite{10.1145/3508352.3549369}, 
a static analysis framework, classifies 96 hardware CWEs into six categories: Non-RTL, 
Functional Simulation, Static Analysis (no context), Static Analysis (with context), 
Static/RTL Elaboration, and Manual Analysis.

Table 1 presents the classification of CWEs under CWE-1194. 

\begin{table*}[t]
\centering
\small
\renewcommand{\arraystretch}{1.2}
\setlength{\tabcolsep}{6pt}

\begin{tabularx}{\textwidth}{|>{\raggedright\arraybackslash}p{0.12\textwidth}
                            |>{\raggedright\arraybackslash}p{0.22\textwidth}
                            |>{\raggedright\arraybackslash}X
                            |>{\centering\arraybackslash}p{0.10\textwidth}|}
\hline
\multicolumn{2}{|l|}{Classification} & CWE-ID & \# of CWEs \\
\hline

\multirow{2}{*}{\parbox[t]{0.12\textwidth}{CWE covered}}
& Convertible to 3-tuple
& 1190, 1193, 1274, 1220, 1222, 1262, 1280, 1299, 1209, 1221, 1223, 1224, 1231, 1232, 
1233, 1234, 226, 1257, 1282, 1310, 1315, 1317, 1279, 1256, 1271, 1314, 1191, 1243, 
1244, 1258, 1272, 1313, 276, 1254, 1239, 1330, 1270
& 37 \\
\cline{2-4}

& Partially Convertible to 3-tuple
& 1264, 1283, 1328, 441, 1189, 1240, 1260, 1268, 1294, 1302, 1334, 1245, 1250, 1253, 
1252, 1251, 1311, 1312, 1316, 1431, 1320, 1296, 1429, 1281, 1259, 1290, 1292
& 27 \\
\hline

\multirow{4}{*}{\parbox[t]{0.12\textwidth}{CWE not covered}}
& Physical Env
& 1247, 1248, 1261, 1278, 1319, 1351, 1384, 1263, 1301, 1304, 1338, 1298, 1266, 1297
& 14 \\
\cline{2-4}

& Side channel / IFP
& 1255, 1300, 1323, 203, 1331, 1420, 1342, 1303, 319, 1421, 1422, 1423
& 12 \\
\cline{2-4}

& SW/Alg/ISA/System Design
& 1332, 1277, 1329, 1357, 1241, 1246, 1318, 1326, 1269
& 9 \\
\cline{2-4}

& Require protocol info/documentation
& 1053, 1059, 440, 1291, 1295, 325, 1192, 1242, 1267, 1276, 1273
& 11 \\
\hline
\end{tabularx}

\caption{Classification of hardware CWEs in our framework.}
\label{tab:cwe_classification}
\end{table*}

\subsection{Capturing the Hierarchy of CWE Entries}

To leverage the hierarchical nature of the CWE entries, we build a tree of CWE entries.
The root of the tree is CWE-1000: Research Concepts. As stated in MITRE's website~\cite{cwe}, 
``[CWE-1000] is mainly organized according to abstractions of behaviors instead of 
how they can be detected, where they appear in code, or when they are introduced 
in the development life cycle.'' This view uses a deep hierarchical 
organization, with more levels of abstraction than other classification schemes.
The large tree has 10 subtrees, one for each of the 10 Pillars in
CWE-1000. These pillars express high-level security intents.

The subtrees are stored in JSON format. For each CWE node inside the subtree, we store its CWE-ID, name, description, mapping status
(whether it is mappable or discouraged by MITRE), and its children nodes. Only CWEs that are mappable and 
covered by our framework can be used as target CWEs for CHARGE. The CWE hierarchy
is built once and is part of the CHARGE framework.  
 
We start from each mappable CWE, trace upward following the "Child Of" relationship, 
building the edges and intermediate nodes until we reach a pillar. In this framework, 
we consider only the Parent–Child relationship. Other relationships in the CWE-1000: 
Research Concepts, such as CanPrecede and CanFollow, are excluded because they do 
not encode strict hierarchical structure and would unnecessarily complicate the
framework.

\begin{figure*}[t]
\centering
\includegraphics[width=\textwidth]{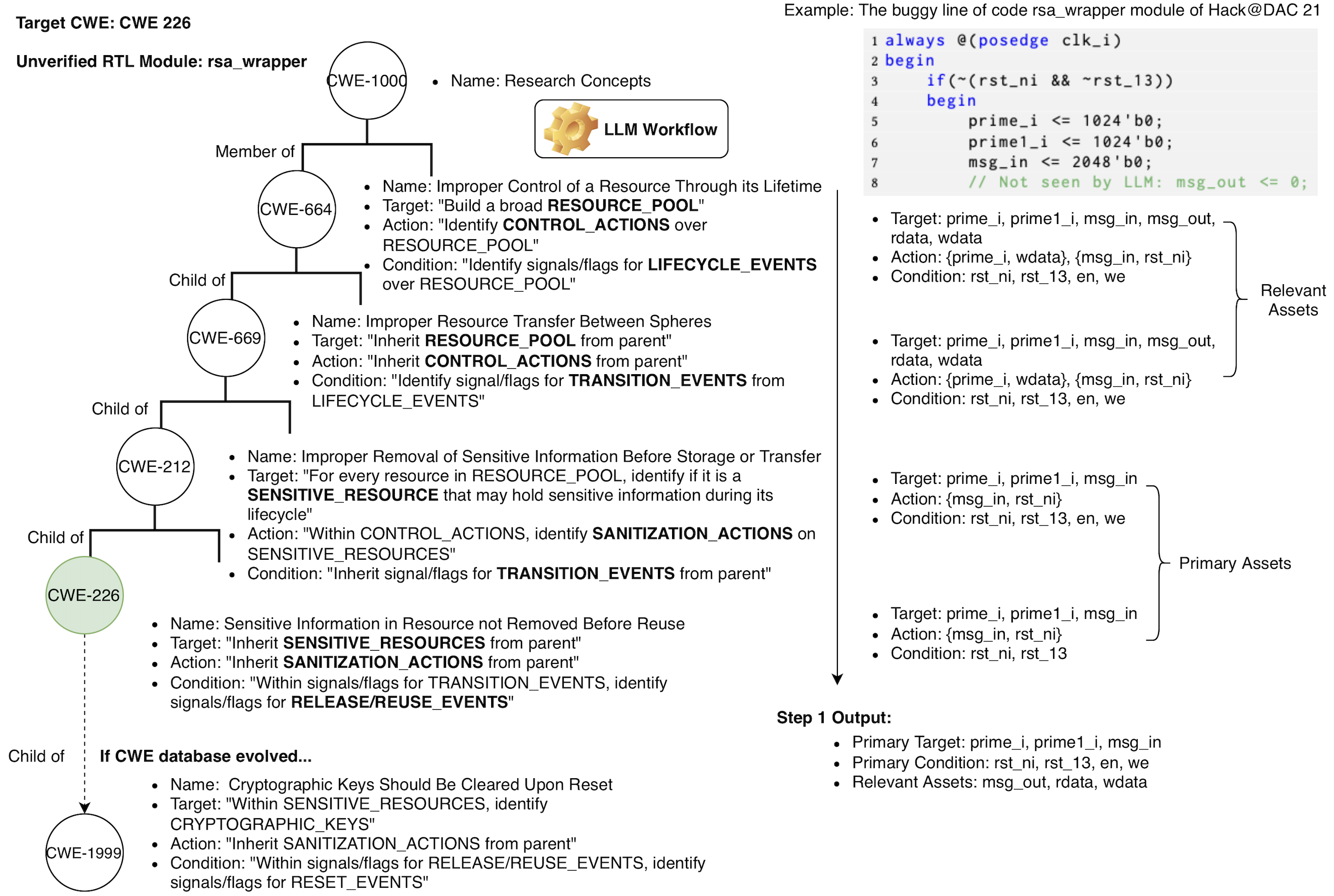}
\caption{Explanation of CHARGE's Usage of CWE Hierarchies}
\label{fig:hierarchy}
\end{figure*}

Each CWE node has an additional field, ``asset identification,'' which includes three subfields (target, action, condition) that are used
to build the prompt strings used for asset identification in step 1 of the
framework (Sec.~\ref{sec:step1}). Figure ~\ref{fig:hierarchy} shows an example of
the sequence of prompt strings for CWE-226, starting from the root node (CWE-664).

Importantly, our methodology assumes that the CWE database is organized hierarchically and that a child 
CWE represents a refinement of its parent. Under this assumption, newly introduced CWEs 
can be incorporated by inheriting and specializing the corresponding 3-tuples of their ancestors.
To illustrate this process, Figure~\ref{fig:hierarchy} includes a hypothetical child CWE, 
CWE-1999 (Cryptographic Keys Should Be Cleared Upon Reset). CWE-1999 inherits the 
sanitization-related security intuition of CWE-226 while specializing the target to 
cryptographic keys and the condition to reset events. Existing CWE definitions and prompts 
remain unchanged, enabling CHARGE to scale with the continued growth of the CWE database 
while preserving backward compatibility.
\section{CHARGE Framework}
\begin{figure*}[t]
\centering
\includegraphics[width=\textwidth]{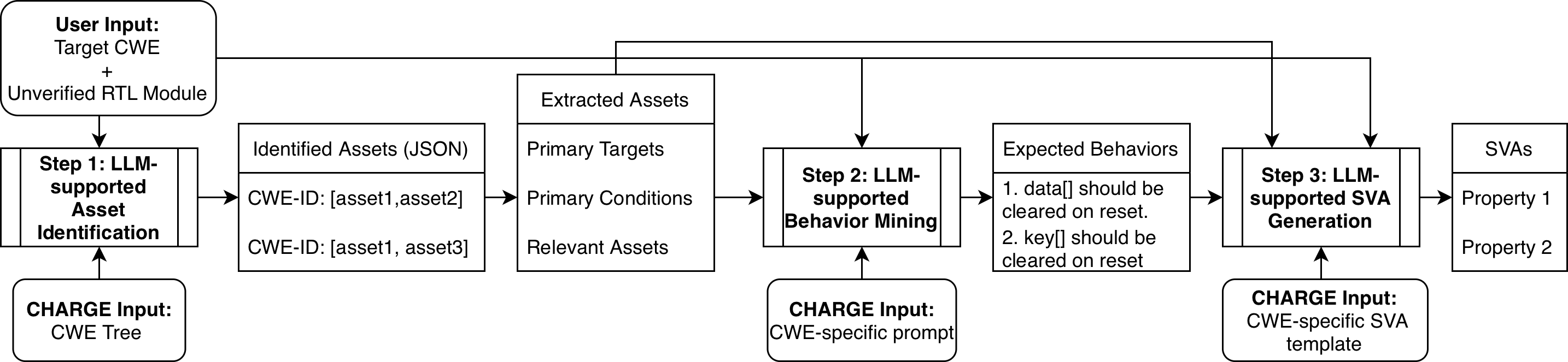}
\caption{Overview of our framework. 
    }
\label{fig:framework}
\end{figure*}

Given a target CWE ID and a RTL module, CHARGE will automatically generate a set of SystemVerilog Assertions (SVAs) 
that check the security policy indicated by the CWE for the
given RTL. CHARGE's workflow is illustrated in Figure~\ref{fig:framework}.
Two aspects of note: we assume the RTL is unverified and potentially buggy, and
we do not require design specifications as input.

The first step of the framework has two jobs: 1. to determine whether the given
CWE is suitable for the framework, and 2. if it is, to output the assets in the
RTL along with the desired behaviors and the conditions under which those
behaviors hold. The output is formatted as a JSON. If the CWE is found not to be relevant, the framework reports
``inapplicable CWE'' in step 1. CHARGE collects the assets from step 1 and put them under 3 categories:
primary targets, primary conditions, and relevant assets. 
The second step takes these three categories of assets from step 1 along with the
RTL and CWE and outputs the expected behavior, stated in natural language. The
third step takes the natural language description from step 2, the three categories of assets from
step 1, the RTL, and CWE and outputs the generated properties. All three steps
use an LLM to generate the desired output.

The generated properties are in standard SystemVerilog Assertion (SVA) format and can be
used with any testing or formal verification tool that accepts the SVA
properties. In our evaluation, we use the Cadence JasperGold 
FPV model checker to check for property violations, but open source verification
frameworks that accept SVA properties could also be used (e.g.,~\cite{symbiyosis,Ryan2025Sylq}).

We describe the three steps of the framework in detail next.

\subsection{Step 1: LLM-supported Asset Identification}
\label{sec:step1}
CHARGE starts with asset identification, which is the most critical step in the framework. 
Given a target CWE, CHARGE traverses the corresponding 
path in the CWE hierarchy, starting from the children of CWE-1000 and proceeding toward the target node.
At each level, the LLM identifies candidate assets.

The target CWE ID is first used by CHARGE to retrieve the corresponding hierarchical path from the 
stored CWE tree. The extracted path, together with the CWE hierarchy represented as a JSON file, is then 
incorporated into the LLM prompt. The prompt instructs the LLM to follow the extracted path
when processing the CWE hierarchy JSON file. The LLM follows the path and processes the queries 
in the 3-tuple structure to identify assets for 
each CWE node. The output is a JSON file that includes the identified assets for each CWE node 
along the path. The user prompt is only processed once by the LLM, and the output is generated 
in one shot.

If the LLM is not able to identify any asset for the target CWE, CHARGE will report the target CWE is not applicable 
to the given RTL module. Otherwise, the identified assets 
for the target CWE and its parent are collected as primary assets and the identified assets for 
other ancestors are collected as relevant assets.

By starting at CWE-1000's children, which expresses the most abstract security intuition, and querying for assets with
each CWE along the path to the target CWE, CHARGE is able to identify assets
relevant to the target even when the (potentially buggy) RTL does not make their
relevance clear. 
For example, consider the buggy code snippet from Hack@DAC 2021 shown on
%Listing~\ref{lst:bug76hackdac21}. 
the top right of Figure~\ref{fig:hierarchy}.
Here the target CWE is CWE-226 "Sensitive Information in 
Resource Not Removed Before Reuse". The line of code that is commented out in
the listing is missing from the RTL and is not seen by the LLM. In other words,
the LLM does not see the comment; we include it here only to make the bug clear.

If we ask the LLM to identify the target assets for CWE-226 without using the
hierarchy of CWEs, the LLM may not identify \texttt{msg\_out} as a 
potential target because the expected behavior of zeroizing \texttt{msg\_out} during reset is missing.
CHARGE's hierarchical-based asset identification mitigates this problem by
collecting asset information 
from the ancestor with highest level of abstraction. Although \texttt{msg\_out}
is not being zeroized when reset, it is a target asset for CWE-664 "Improper
Control of a Resource Through its Lifetime," the great-grandparent of
CWE-226. Because \texttt{msg\_out} is read and written in the module (not shown
in the listing), the LLM identifies it as a relevant asset for CWE-664.
Figure~\ref{fig:hierarchy} illustrated this process on the right side. 
We evaluate the effectiveness of this approach in the evaluation (Section~\ref{sec:baseline1}).

\subsection{Step 2: LLM-supported Behavioral Mining}

The assets extracted from Step 1's output are then used as the inputs for mining expected security behaviors.
Other inputs include unverified RTL code, the target CWE's description, and a manually crafted policy 
question based on the CWE's security intent. The motivation for this step is to
prompt the LLM to generate a list of expected security behaviors using the CWE
rather than the possibly buggy RTL.

To illustrate this step, we will continue to use the same code snippet and target CWE in Step 1. We take 
the primary targets, primary conditions, and relevant assets identified in Step 1 as the inputs. In this example, 
the primary targets are \texttt{prime\_i}, \texttt{prime1\_i}, and \texttt{msg\_in}; the primary conditions are \~rst\_ni, rst13, \texttt{en}, and \texttt{we}; the 
relevant assets are \texttt{msg\_out},\texttt{rdata}, \texttt{wdata} . Since action expresses a behavior, we will only extract the assets 
involved in the action as part of the primary or relevant assets. 

To bridge the gap between security-critical assets and expected security behaviors, we manually craft a policy 
question based on the CWE's security intent, which serves as the "action". For example, for CWE-226, the question is 
"under what condition should the given asset(s) be zeroized/cleared?"
Each question is written once per CWE and is reused across target RTL modules. Using the identified assets, 
target CWE and prompt, and the unverified RTL code as context, the LLM generates a list of 
expected security behaviors. For example, "when reset is on, \texttt{msg\_out} needs to be cleared," "when reset is on, 
\texttt{prime\_i} needs to be cleared." These expected behaviors can then be
used to generate SVAs in the next step.

We stored each mappable CWE's ID, name, description, and the prompting question all in one JSON file. When running this step, 
only the target CWE's information is used. The output of this step is a list of expected security behaviors 
in natural language, stored in a .txt file. CHARGE will later use this list to generate SVAs in Step 3.
 
In the user prompt, we explicitly ask the LLM to answer the prompting question and generate expected 
behaviors for the target CWE, as well as only use the unverified RTL code as a contextual reference. This is to avoid the 
issue of the LLM incorrectly assuming the behavior of the unverified RTL code. However, in some cases, the bug inside 
the unverified RTL code might still get baked into the expected behavior (see Section~\ref{sec:misledrtl}).

\subsection{Step 3: LLM-supported SVA generation}

The list of expected behaviors generated in Step 2 are then used to generate SVAs, along with the extracted assets from Step 1's output, target CWE, 
unverified RTL code, and a manually crafted SVA template based on the CWE semantics. All SVA templates are stored in a JSON file, where for each mappable CWE, 
we include the CWE ID, name, description, and the SVA template. When running this step, only the target CWE's information is used.
Similar to the CWE tree and the prompting questions, the SVA templates are
created once per CWE and reused across different RTL modules. The templates are
designed to be JasperGold inline assertions. See, for example, Listing~\ref{lst:sva}.

The LLM is prompted to generate a set of SVAs by filling in the template based on the expected behaviors and other inputs. 
There's no limit to the number of SVAs generated for each expected behavior, but each generated SVA should be mapped to at least one expected behavior. 
LLM is explicitly asked to only use the real RTL signals when generating SVAs, and should neither invent new signals nor use syntax that is not supported by 
JasperGold (e.g. using for loops to generate multiple SVAs). The output of this step is in a .sv file that includes all generated SVAs. 
To later evaluate vacuity, we also ask the LLM to generate a cover property for each generated SVA, which is in the format of "cover -name CWE226\_cover {(reset\_trigger\_event)}", and is provided to LLM as part of the SVA template. 
This cover property is used to check if the assertion is vacuously true. 
After this step, we will have a set of SVAs that can be used to check the security property represented by the target CWE.

\section{Implementation}
We use OpenAI's GPT-4.1 API for all LLM interactions. We set the temperature to
zero to reduce the nondeterminism of LLM outputs. 

CHARGE is fully automated, with the necessary information automatically extracted from each step's output and passed to the next step.
The implementation of CHARGE consists of a single Python script, in which each step is implemented as a separate function along with 
several helper functions. The framework writes the output of each step to a separate file and logs all LLM interactions.
CHARGE is available on GitHub: https://github.com/HWSec-UNC/CHARGE.

\section{Evaluation}

To evaluate CHARGE, we compare to the manually written properties for the three
Hack@DAC designs in the public Verification Benchmarks
repository~\cite{rogers2025hardware}. The designs are a PULPissimo SoC (Hack@DAC
2018), a CVA6 SoC (Hack@DAC 2019), and an OpenPiton SoC (Hack@DAC 2021). The
repository provides properties covering 20, 11, and 17 bugs for the three designs,
respectively. Of those, the repository has marked one, one, and two properties,
respectively, as ``no violation
found,'' meaning the given property does not catch the relevant bug in the
design. We exclude these four cases from our evaluation, as they do not 
provide a reliable ground-truth reference. We further exclude two additional 
cases. The first is a clock glitching bug in Hack@DAC 2018, which falls outside our 
scope as it requires explicit reasoning about clock frequency. The second is a hardcoded 
HMAC key in Hack@DAC 2021. We found that the corresponding property in the repository is incorrect, 
and upon manually inspecting the RTL, we were unable to manually write a property to 
catch this bug that wasn't trivially tailored to the specific hardcoded
value. We noted this issue in a pull request to the repository.

After excluding these six bugs, our end-to-end evaluation (Sec.~\ref{sec:endtoend}), which focuses on 
bug detection, covers 25 RTL modules and 42 distinct security bugs. These 
bugs map to 20 unique CWEs, which share 6 common pillars in the CWE hierarchy. 
Our baseline evaluation (Sec.~\ref{sec:baseline}) focuses on Hack@DAC 2021, and is designed to demonstrate 
the effectiveness of our overall framework as well as its components across different LLMs. Using LLM generated assertions, 
we are able to find 3 errors for the assertions in Verification Benchmark repository, and 1 undocumented 
bug in Hack@DAC 2021's OpenPiton SoC.

\subsection{End-to-End Bug Detection Results}
\label{sec:endtoend}
We evaluate CHARGE's performance on end-to-end bug detection by checking if 
the generated assertions can detect the known bugs inside the Verification Benchmarks.
All the assertions generated by CHARGE are JasperGold inline assertions, 
and we run Cadence JasperGold FPV to check if the generated assertions can detect the known bugs.
We made the following edits to the generated assertions in order to use them in JasperGold:
deleting the posedge clock keywords (the designs are set up with default clocks 
in JasperGold environment), deleting the semicolon at the end of the assertion
if it exists, and adding a pair of brackets around the assertion body if needed. 

To account for the nondeterminism of LLM outputs, we run CHARGE three times on
each target CWE--RTL module pair.
We summarize the bug detection results in Table~\ref{tab:bug_eval}. 
CHARGE generates useful assets for 34 out of 42 known bugs, generates meaningful expected behaviors 
for 29 out of 42 known bugs, and generates assertions that detect 27 out of the 42 known bugs.

The ``Asset'' column in Table~\ref{tab:bug_eval} is checked if all the necessary
assets are named in the generated property. In other words, if one of
the following three conditions holds: 1) all the assets
used in the Verification Benchmark's property's body
are in the generated assets; 2) the generated expected behaviors
(``Behavior'' column) will include the missing asset, and the generated
behavior is meaningful for assertion generation; or 3) the generated assertions detect the bug
without the missing asset. Otherwise, the column is marked with an ``x''.
To understand the second case, consider Hack@DAC 2021 P39. The bug is "AES plain text is left uncleared after the encryption is over in the peripheral registers". 
The LLM does not identify \texttt{ct\_valid} (ciphertext valid) as the core
asset after step 1. However, the LLM does identify \texttt{ct}
(ciphertext) and \texttt{pc} (plaintext) as primary targets for CWE-226, and
therefore, in step 2 produces the correct behavior that plaintext should be
cleared when encryption is over. We neither identify nor input
\texttt{ct\_valid} into step 2, but LLM is able to produce it in its output.

We determine whether a generated behavior is meaningful for assertion generation
through manual inspection of the RTL and ``Bug Description'' in the Verification
Benchmark. Meaningful behavior is in some sense the opposite of the buggy behavior.
To check whether the generated assertion can detect the bug, we run JasperGold
FPV with the generated assertion, and check whether the assertion is violated by
the buggy design, and if so, then either the generated assertion must be semantically equivalent to the 
manually written property in the Verification Benchmarks that detects the same
bug; or we determine through manual inspection of the RTL that the violation is
detecting the same bug.

\begin{table}[t]
\centering
\scriptsize
\setlength{\tabcolsep}{3pt}
\begin{tabular}{l l l l c c c}
\toprule
\textbf{Design} & \textbf{Module} & \textbf{CWE} & \textbf{Bug} & \textbf{Assertion} & \textbf{Behavior} & \textbf{Asset} \\
\midrule

\multirow{18}{*}{Hack@DAC18}
& mux\_func & CWE-226 & 29 & \cmark &  \cmark & \cmark \\
& mux\_func & CWE-1240 & 21 & \cmark & \cmark & \cmark \\

& riscv\_cs\_registers & CWE-1220 & 27 & \xmark & \xmark & \cmark \\
& riscv\_cs\_registers & CWE-1189 & 3 & \xmark & \xmark & \xmark \\

& adbg\_tap\_top & CWE-1221 & 16 & \cmark & \cmark & \cmark \\
& adbg\_tap\_top & CWE-1244 & 12 & \xmark & \xmark & \xmark \\
& adbg\_tap\_top & CWE-1254 & 9 & \cmark & \cmark & \cmark \\
& adbg\_tap\_top & CWE-1254 & 10 & \cmark & \cmark & \cmark \\

& riscv\_alu & CWE-1245 & 14 & \xmark & \cmark & \cmark \\
& riscv\_controller & CWE-1245 & 13 & \cmark & \cmark & \cmark \\
& riscv\_debug\_unit & CWE-1262 & 11 & \cmark & \cmark & \cmark \\

& periph\_bus\_defines & CWE-1260 & 8 & \cmark & \cmark & \cmark \\
& periph\_bus\_defines & CWE-1260 & 1 & \cmark & \cmark & \cmark \\
& periph\_bus\_defines & CWE-1257 & 6 & \cmark & \cmark & \cmark \\

& axi\_address\_decoder\_AR & CWE-1245 & 7 & \cmark & \cmark & \cmark \\
& apb\_gpio & CWE-1189 & 5 & \xmark & \xmark & \cmark \\
& apb\_gpio & CWE-1231 & 4 & \cmark & \cmark & \cmark \\
& soc\_interconnect & CWE-1189 & 2 & \cmark & \cmark & \cmark \\

\midrule

\multirow{6}{*}{Hack@DAC19}
& axi\_node\_intf\_wrap & CWE-1220 & 1 & \cmark & \cmark & \cmark \\

& csr\_regfile & CWE-1262 & 9 & \xmark & \cmark & \cmark \\
& csr\_regfile & CWE-1220 & 24 & \xmark & \xmark & \xmark \\
& csr\_regfile & CWE-1220 & 25 & \xmark & \xmark & \xmark \\
& csr\_regfile & CWE-1220 & 29 & \xmark & \xmark & \xmark \\

& commit\_stage & CWE-1281 & 21 & \xmark & \xmark & \cmark \\
& commit\_stage & CWE-1281 & 22 & \cmark & \cmark & \cmark \\

& ariane & CWE-1281 & 23 & \xmark & \xmark & \xmark \\
& ariane & CWE-1281 & 26 & \xmark & \xmark & \cmark \\

& controller & CWE-1281 & 32 & \cmark & \cmark & \cmark \\

\midrule

\multirow{11}{*}{Hack@DAC21}
& sha256\_wrapper & CWE-1239 & P36 & \cmark & \cmark & \cmark \\

& aes0\_wrapper & CWE-226 & P39 & \cmark & \cmark & \cmark \\
& aes0\_wrapper & CWE-1258 & P46 & \cmark & \cmark & \cmark \\
& aes0\_wrapper & CWE-1258 & P47 & \cmark & \cmark & \cmark \\

& dmi\_jtag & CWE-1245 & P2 & \cmark & \cmark & \cmark \\
& dmi\_jtag & CWE-1245 & P84 & \xmark & \xmark & \cmark \\

& csr\_regfile & CWE-1262 & P18 & \xmark & \xmark & \xmark \\

& reglk\_wrapper & CWE-1234 & P48 & \cmark & \cmark & \cmark \\
& reglk\_wrapper & CWE-1232 & P35 & \cmark & \cmark & \cmark \\

& rsa\_wrapper & CWE-226 & P95 & \cmark & \cmark & \cmark \\
& acct\_wrapper & CWE-276 & P42 & \cmark & \cmark & \cmark \\
& dma & CWE-1245 & P57 & \cmark & \cmark & \cmark \\
& riscv\_peripheral & CWE-1310 & P96 & \xmark & \xmark & \xmark \\
& aes\_192 & CWE-1240 & P14 & \cmark & \cmark & \cmark \\

\bottomrule
\end{tabular}
\caption{Bug detection results across Hack@DAC Designs}
\label{tab:bug_eval}
\end{table}

\subsection{Comparing to a Baseline}
\label{sec:baseline}
In these experiments we evaluate the effectiveness of each of the features of
CHARGE. The experiments are all conducted on the Hack@DAC 2021 OpenPiton SoC design, which has 14 known security bugs,
covering 10 unique CWEs and 10 unique RTL modules. As before 
we run each configuration three times per target CWE--RTL module pair. 
Moreover, to evaluate the cross-LLM generalization capability, we run CHARGE using three 
models for each Baseline experiment: GPT-4.1, Gemini 2.5 Pro, and Claude Sonnet 4.6.

\subsubsection{Baseline 0}
In this experiment, we prompt the LLM to generate assertions to verify the given
CWE for the given unverified RTL module. The prompt includes explicit
instructions to use real RTL signals. We evaluate the generated assertions on 
three aspects: syntax correctness, reachability, and bug detection
result. Table~\ref{tab:baseline0_cross_llm} shows the results.

\begin{table}[t]
\centering
\small
\caption{Cross-LLM evaluation of Baseline 0 and CHARGE.}
\label{tab:baseline0_cross_llm}
\begin{tabular}{llccc}
\hline
Metric & Method & GPT-4.1 & Gemini 2.5 Pro & Claude Sonnet 4.6 \\
\hline

\multirow{2}{*}{Bug Detection}
& Baseline 0
& 1/14 
& 6/14 
& 3/14  \\
& \textbf{CHARGE}
& 11/14 
& 11/14 
& 11/14  \\

\hline

\multirow{2}{*}{Compilable}
& Baseline 0
& 91.1\%
& 91.5\%
& 97.1\% \\
& \textbf{CHARGE}
& 89.0\%
& 97.0\%
& 97.4\% \\

\hline

\multirow{2}{*}{Unreachable}
& Baseline 0
& 0\%
& 9.3\%
& 2.7\% \\
& \textbf{CHARGE}
& 7.8\%
& 5.9\%
& 9.3\% \\

\hline
\end{tabular}
\end{table}

\subsubsection{Baseline 1}
\label{sec:baseline1}
In this experiment, we evaluate the effect of using CWE hierarchy and 3-tuple 
structure on asset identification. We compare three configurations in Step 1 of CHARGE's framework: 
1. Hierarchy + 3-Tuple: the standard CHARGE workflow for asset identification, which leverages both 
the CWE hierarchy and the 3-tuple structure; 2. No hierarchy + 3 Tuple:
explicitly prompting the LLM to 
only identify assets that are relevant to the target CWE, based on the <target, action, condition> 
structure; 3. No hierarchy + No 3 Tuple: explicitly prompting LLM to only identify assets 
relevant to the target CWE, and not leveraging the 3-tuple structure.

We run each configuration three times per target CWE–RTL module pair, and we introduce a scoring system to 
evaluate the quality of the generated assets. For each bug, we first check the number 
of core assets that are used in the manually written properties in Verification Benchmarks. 
One asset corresponds to one point. Then we check how many of those core assets can be found in the generated 
assets across the three configurations. To account for the nondeterminism of LLM outputs, 
if an asset can be found in all 3 runs, then we mark that asset as "found" and give 1 point; 
if an asset can be found in 1 or 2 out of 3 runs, then we mark that asset as "partially found" and give 0.5 points; 
if an asset cannot be found in any of the three runs, then we mark that asset as "not found" and give 0 points.
Table~\ref{tab:baseline1_cross_llm}, summarizes the performance of each configuration on asset identification.

\begin{table}[t]
\centering
\caption{Baseline 1: Asset identification accuracy across different LLMs.}
\begin{tabular}{lccc}
\hline
Model & Hierarchy+3Tuple & 3Tuple & None \\
\hline
GPT-4.1           & 30/41 & 21.5/41 & 18.5/41 \\
Gemini 2.5 Pro    & 37.5/41 & 28/41 & 24.5/41 \\
Claude Sonnet 4.6 & 38/41 & 35.5/41 & 32/41 \\
\hline
\end{tabular}
\label{tab:baseline1_cross_llm}
\end{table}

\subsubsection*{Summary}
Tables~\ref{tab:baseline0_cross_llm} and ~\ref{tab:baseline1_cross_llm} show that 
CHARGE consistently improves over the corresponding baselines in both bug detection 
and asset identification across all three models. These results suggest that the 
gains primarily arise from CHARGE's hierarchical reasoning and structured decomposition 
rather than from a particular LLM. Detailed per-benchmark results are provided in the Appendix.

\subsection{Finding Errors in the Manually Written Properties}

During the course of our evaluation, we found that three of the manually written properties 
in the public Verification Benchmarks repository were incorrect and CHARGE's generated properties were correct. 
In the Hack@DAC 2018 P29 mux\_func module and Hack@DAC 2021 P95 rsa\_wrapper module, the manually written 
properties used the overlapped implication operator, while the correct operator should be non-overlapped 
implication. In the Hack@DAC 2018 P8 periph\_bus\_defines module, the manually written property specified the
wrong address range for checking address overlap, while CHARGE's generated property specified the correct address 
range. We reported these issues to the public repository, and our pull requests have been accepted. 

\subsection{Failure Mode Analysis}

CHARGE is evaluated on 42 known bugs, and 15 of them are not detected by the generated assertions. 
We analyze the failure modes of these 15 cases, and discuss the limitations of our framework. 

\subsubsection{Commonality of failure cases}

Our evaluation covered 20 unique CWEs. The 15 undetected bugs are mapped to 7 unique CWEs: 4 bugs are
mapped to CWE-1220, 2 bugs are mapped to CWE-1245, 2 bugs are mapped to CWE-1189, 2 bugs are mapped to 
CWE-1262, 3 bugs are mapped to CWE-1281, 1 bug is mapped to CWE-1244, and 1 bug is mapped to CWE-1310.
These 7 CWEs share 4 common ancestors in the CWE hierarchy: CWE-284 (3 children), CWE-710 (2 children),
CWE-664 (1 child), and CWE-691 (1 child). This result suggests that the failure cases are not randomly 
distributed across target CWEs.
The LASHED paper~\cite{ahmad2025lashedllmsstatichardware} also found ``a
significant variation [in]  success based on the CWE.'' 
Although the troublesome target CWEs are mostly different for the two
frameworks, these results suggest that LLM-based assertion generation is
sensitive to choice of CWE. 

Our findings also suggest that the length of the RTL module is a factor
affecting performance. The 27 bugs CHARGE is able to detect originated from 19 different modules 
with an average of 355 lines per module. The 15 bugs CHARGE is not able to detect originated from 
10 different modules with an average of 872 lines per module. Furthermore, 11
of the 15 undetected bugs are from 6 modules with more than 800 lines of code
each. 

\subsubsection{Type1:  Missing Asset/Behavior}

We further classified the 15 undetected cases into two categories. The first
is failure due to the LLM missing the asset or behavior. The bugs in this
category are Hack@DAC 2018 P3, 27, 12, 14; Hack@DAC 2019 P21, 24, 25, 29, 23,
26; Hack@DAC 2021 P18, 96. We identify four root causes:
\begin{itemize}
  \item CWE Mismatch: If the given target CWE does not fit the looked-for bug,
    the LLM will fail to identify the relevant assets for the bug. We found two
    instances where this was the case in the Verification Benchmarks repository.
    
  \item Lack of evidence in RTL: If the core asset is completely missing from the RTL, 
  or the evidence is too weak, the LLM will not be able to identify that asset. 
  \item Larger RTL modules and more complex behaviors: This category is mostly overlapped with 
  the 11 undetected bugs that have an average of 1180 lines of code per module. For example, 5 failure cases are from the csr\_regfile module, which has more than 1000 lines of code.%% , 
   
  \item The expected behavior is not within the training data of the LLM: In
    some cases, the LLM was able to identify the assets, but was not able to
    build a connection between these assets to generate the expected behavior 
  that can enforce the security property represented by the target CWE. We suspect that this is because the 
  expected behavior between these assets is not within the training data of the LLM.
\end{itemize}

\subsubsection{Type2: Behavior/Assertion Generation is Misled by the RTL}
\label{sec:misledrtl}
In the second category of failure cases, the LLM is able to build a connection
between the identified assets, but the connection is wrong (often the
opposite). The bugs in this category are Hack@DAC 2018 P5, Hack@DAC 2019 P9, and Hack@DAC 2021
P84. We suspect the LLM is misled by the buggy RTL.  
For example, in Hack@DAC 2021 P84, the target CWE is CWE-1245 Improper FSM. The
expected state transition is missing from the RTL, so the LLM excludes the
state from the valid next states.

\subsection{Case Study: Unknown Security Weakness Detection}

CHARGE generated an assertion for CWE-226 
in aes0\_wrapper module that is violated by the design, but not on the HACK@DAC21 bug list.
The generated assertion states that key material should be cleared on reset
and is given in Listing~\ref{lst:resetkeys}.

\begin{lstlisting}[
    caption={The assertion detecting an unknown security flaw},
    label={lst:resetkeys},
    float={tbph},
    style={verilog-style}]
  assert -name reset_p1 {
  (!aes0_wrapper_i.rst_ni || aes0_wrapper_i.rst_1) |=>
    (aes0_wrapper_i.key0 == '0)
    }
\end{lstlisting}

Using JasperGold FPV, we constructed a trace as a cover property
showing that reset occurs, no key write happens after reset, but AES
encryption is initiated, the key used is non-zero and originates from pre-reset
state, and the key is actively used in the AES computation. The trace is
reachable and indicates that the key material is retained and reused across
reset. This behavior violates a standard security property: cryptographic keys should be cleared upon reset. 
We reported this issue to the Hack@DAC 2021 repository.

\subsection{Bug-to-CWE Mapping Algorithm leveraging CWE Hierarchy}

In the Verification Benchmarks repository some bugs have no assigned CWE while
others have CWEs that is marked as "Discouraged" on Mitre's CWE website. To
address this issue, we leverage the structure of the CWE hierarchy to help with bug-to-CWE mapping.
We are inspired by existing research in software security using the CWE
hierarchy with an LLM to map CVEs to CWEs or Bugs to CWEs using Github's open
source bug-fix commits~\cite{10172785}.

We use a greedy algorithm for bug-to-CWE mapping, leveraging the
CWE-hierarchies. We start from the root node of the CWE hierarchy, and
ask the LLM to select the most suitable child node based on the bug description
and code snippet, iterating until a mappable CWE is found. Using this algorithm,
we successfully mapped 7 bugs previously mapped to a Discouraged CWE in
Verification Benchmarks to a mappable CWE. CHARGE successfully identifies the core assets using 
the newly mapped CWEs. Further research can be done to improve the preciseness
by fine-tuning a model instead of using  
commercial LLMs, or improving the algorithm itself.
\section{Related Work}

\subsubsection*{Using LLMs to Generate Properties}

Current work on using LLMs to generate SVA can be categorized across two
axes: using commercial 
LLMs with prompt engineering vs. using customized LLMs, and generating security
SVA vs. generating functional SVA. 
For functional SVA generation, design specification is usually provided as a ground truth, 
while for security SVA generation, there's more variation in the input context, as there is no formal 
specification for security properties of a design. Examples of frameworks for generating functional SVAs not using customized LLMs include 
FLAG~\cite{shih2025flag} and ChIRAAG~\cite{mali2024chiraag}. AssertLLM~\cite{fang2026assert} uses customized LLMs, 
and VERT~\cite{10.1145/3764934} presents an open-source dataset that can help
with fine-tuning LLMs for SVA generation. 
Kande et al.~\cite{kande2024security} are one of the first to explore 
the use of LLMs to generate security SVA. SVAgent~\cite{guo2025svagent} uses
prompt engineering to break down the security SVA generation process into subquestions. Meng et al.~\cite{meng2023unlockinghardwaresecurityassurance} proposed an automated NLP-based security property generator. 

\subsubsection*{Using CWEs to Generate or Validate Properties}

% The DIVAS paper presents a framework in which the user provides details about a design and
% its security requirements, and LLMs are used to identify the relevant CWEs and generate assertions, 
% from which a tuple describing the SVA's predicate, timing, and action can be extracted and used 
% by existing tools to generate compliant Verilog RTL~\cite{paria2023divas,paria2024spell}. 
% However, the authors report many shortcomings in the generated SVAs, such as invalid
% syntax, invalid keywords, and incorrect logic.

CWEAT~\cite{10.1145/3508352.3549369} uses CWEs to guide static analysis of 
early-stage hardware designs by building scanners for RTL, each designed to detect the potential weaknesses associated with a specific CWE. 
Other work includes DIVAS ~\cite{paria2023divas}, which uses LLMs to identify
relevant CWEs and generate security SVA;
LASHED~\cite{ahmad2025lashedllmsstatichardware}, which uses the LLM to identify
assets for 5 different CWEs, and then uses static analysis to produce SVAs; and
SoCureLLM ~\cite{cryptoeprint:2024/983}, which uses CWEs to represent the set of
vulnerabilities, and divides a large codebase into smaller parts to delegate to different LLMs. 

\subsubsection*{Mining Design Behavior to Generate Properties}
GoldMine ~\cite{goldmine2010} automatically generates assertions by 
mining data from simulation traces and using static analysis of RTL designs. 
Isadora ~\cite{10.1145/3474376.3487286} focuses on mining information flow properties from RTL designs. 
Deutschbein et al. ~\cite{9300291} evaluates security specification mining for a CISC architecture, and targets properties written at the ISA level.

\subsubsection*{Automatically Translating Properties}
Transys ~\cite{9152775} translates security properties written for one hardware design to 
analogous properties suitable for a second design. Transys works on both trace property and informational flow 
property, and the translation is not using LLMs. 

\subsubsection*{Writing Properties Manually}
Rogers et al. ~\cite{rogers2025hardware} developed 120 SystemVerilog Assertions for four open-source designs, 
and they are available in the Verification Benchmark repository. We use these assertions as a benchmark 
to evaluate the effectiveness of CHARGE.

\section{Discussion}
\subsection{Positioning of CHARGE}

\begin{table*}[t]
\centering
\caption{Recent LLM-based frameworks for hardware security verification and analysis.}
\label{tab:security_frameworks}
\small
\begin{tabular}{p{1.5cm}p{3.3cm}p{3cm}p{2cm}p{6.3cm}}
\hline
Framework & Required User Input & Final Output & Open Source Tool & Evaluation Benchmarks \\
\hline
Assertain ~\cite{tarek2026assertainautomatedsecurityassertion} & RTL + Threat Model & SVAs & \xmark & 11 RTL designs \\
LAsset ~\cite{hasan2026lassetllmassistedsecurityasset} & RTL (+ optional Spec.) & Security Asset List & \xmark & NEORV32, OpenTitan/OpenCores \\
LASSO ~\cite{11189178} & RTL + Specification + Documentation & SVAs, Bugs, Coverage report & \cmark & MIT-CEP, OpenTitan, Hack@DAC24 \\
LASHED ~\cite{ahmad2025lashedllmsstatichardware} & RTL + Target CWE  & Bug explanation and localization & \xmark & Hack@DAC21, OpenTitan, e203, Veerwolf \\
SVAgent ~\cite{guo2025svagent} & RTL + Threat Model + Security requirement & SVAs & \xmark & TrustHub, HOST, PyVerilog \\
SoCureLLM ~\cite{cryptoeprint:2024/983} & RTL & Violation report, Security policies & \xmark & Hack@DAC18, TrustHub, HOST22, CVA6, Ibex, CV32E40P\\
\textbf{CHARGE}& RTL + CWE ID & SVAs & \cmark & 25 RTL modules from Hack@DAC18/19/21 \\
\hline
\end{tabular}
\end{table*}

Table~\ref{tab:security_frameworks} summarizes recent LLM-based hardware security frameworks. 
Existing approaches span different stages of the verification pipeline, including asset identification, 
vulnerability classification, policy generation, bug localization, and SVA generation. Consequently, 
they require different user inputs, produce different verification artifacts, and are evaluated on 
different benchmark suites. These frameworks address complementary aspects of hardware security verification.

\begin{table*}[t]
\centering
\caption{Comparison of how recent frameworks utilize CWE knowledge.}
\label{tab:cwe_comparison}
\small
\begin{tabular}{p{1.5cm}p{8cm}c}
\hline
Framework & Highlighted CWE Usage & Systematic CWE Reasoning Framework \\
\hline
CWEAT ~\cite{10.1145/3508352.3549369} &
5 CWE-specific static-analysis scanners. &
\xmark \\
DIVAS ~\cite{paria2023divas} &
Generates security properties using CWEs derived from design specifications &
\xmark \\
SVAgent ~\cite{guo2025svagent}&
Uses CWEs as predefined threat models to decompose security requirements into sub-questions. &
\xmark \\
LASHED ~\cite{ahmad2025lashedllmsstatichardware}&
CWE-specific asset identification and analysis (5 CWEs). &
\xmark \\
SoCureLLM ~\cite{cryptoeprint:2024/983}&
CWE-inspired threat models and security policies. &
\xmark \\
LAsset ~\cite{hasan2026lassetllmassistedsecurityasset}&
Maps identified assets to relevant CWEs for validation. &
\xmark \\
Assertain ~\cite{tarek2026assertainautomatedsecurityassertion}&
Maps RTL and threat model to CWEs, then generates properties for the intersection of the two CWE sets.&
\xmark \\
\textbf{CHARGE} &
Uses CWE hierarchies, 3-tuples, and assertion templates to derive assets, security intent, and generate SVAs from RTL. &
\cmark \\
\hline
\end{tabular}
\end{table*}

As shown in Table~\ref{tab:cwe_comparison}, recent frameworks have explored a variety of 
ways to incorporate CWE knowledge, including vulnerability classification, threat modeling, 
asset identification, and property generation. Together, these approaches demonstrate the 
value of CWE as a source of security knowledge for hardware verification.

CHARGE differs from these approaches in that it treats the CWE taxonomy as a structured 
knowledge source rather than using individual CWE descriptions in isolation. By combining 
hierarchical relationships, a 3-tuple abstraction, and assertion templates, CHARGE provides a 
systematic and reusable methodology for transforming CWE knowledge into security verification 
artifacts, including assets, security intent, and SVAs.
This approach also enables newly introduced CWEs to be incorporated by inheriting and 
refining the security knowledge of their ancestors, allowing CHARGE to evolve naturally with the CWE taxonomy.

Existing approaches such as AST-based representations, retrieval-augmented generation, 
iterative refinement, and agent-based reasoning are largely orthogonal to CHARGE.
These techniques improve the quality of assertion generation once the desired security 
intent is available, whereas CHARGE focuses on systematically deriving that security intent 
from RTL and structured CWE knowledge.

%%%%%%%%%%%%%%%%%%%%%%%%%%%%%%%%%%%%%%%%%%%%%%%%%%
\subsection{Technical Considerations}
%%%%%%%%%%%%%%%%%%%%%%%%%%%%%%%%%%%%%%%%%%%%%%%%%%%

\textbf{Runtime.}
CHARGE typically requires less than one minute to generate 
SVAs for a given RTL module and target CWE, including three LLM calls. Runtime is 
primarily determined by RTL size and API latency. Property validation in Cadence 
JasperGold typically requires less than one minute per run to either prove the property or generate a counterexample.

\noindent
\textbf{Toolchain Dependencies.}
CHARGE is compatible with Python 3.10+ and currently relies on commercial LLM 
APIs together with the Cadence JasperGold formal verification tool. Aside from 
these dependencies, the framework itself is fully automated, with intermediate 
outputs generated and propagated between stages without manual intervention.

\noindent
\textbf{Portability.}
CHARGE generates standard SystemVerilog Assertions (SVAs) and is therefore applicable 
to verification environments that support SVA. In the current implementation, the 
final prompting stage includes JasperGold-specific constraints to avoid unsupported 
language constructs such as \texttt{for} loops and \texttt{genvar}. Adapting CHARGE 
to another SVA-compatible verification environment therefore primarily requires 
modifying the final prompting stage, while the identified assets and inferred security 
behaviors remain unchanged.

%%%%%%%%%%%%%%%%%%%%%%%%%%%%%%%%%%%%%%%%%%%%%%%%%%
\subsection{Scope and Limitations of the Evaluation}
%%%%%%%%%%%%%%%%%%%%%%%%%%%%%%%%%%%%%%%%%%%%%%%%%%%

We evaluate CHARGE using the Hack@DAC designs together with the reference properties provided by 
the open-source Verification Benchmarks repository ~\cite{rogers2025hardware}. The Hack@DAC designs provide buggy RTL implementations, 
while the Verification Benchmarks repository provides manually written security properties that serve 
as ground truth. These reference properties enable comparison of identified assets, inferred expected security 
behaviors, and bug-detection capability. 

Our evaluation is limited to benchmark cases for which both buggy RTL implementations 
and reference properties are available within the selected Hack@DAC designs. 
Consequently, although 
CHARGE supports a broader set of hardware CWEs, not all supported CWEs are represented in 
the current benchmark suite. Extending the evaluation to 
additional benchmark suites and hardware designs is an important direction for future work 
and would enable assessment of a broader set of CWEs.
\section{Conclusion}
We have presented CHARGE, an LLM-based automated framework for generating security
SVAs for unverified RTL designs. CHARGE leverages the
hierarchical nature of CWE entries to guide the LLM toward more accurate and
reliable asset identification. In our evaluation, CHARGE produces SVAs that
catch 27 of 42 bugs in recent Hack@DAC designs. When
comparing to an open-source set of manually written properties, CHARGE
identifies three errors in the set, producing the correct SVAs. Furthermore, CHARGE
generates an SVA that identifies a previously unreported bug in the Hack@DAC 2021 design.  

\section*{Acknowledgments}
We are grateful for the feedback we received from anonymous reviewers; 
the paper benefited from their insights and questions. 
This paper reports on work supported by the National Science
Foundation under Grant No. CNS2247754 and by Intel under the
Scalable Assurance program. This research was also funded in part by a
Summer Undergraduate Research Fellowship from the Office for
Undergraduate Research at the University of North Carolina at
Chapel Hill.

\clearpage

%% %%
%% %% The acknowledgments section is defined using the "acks" environment
%% %% (and NOT an unnumbered section). This ensures the proper
%% %% identification of the section in the article metadata, and the
%% %% consistent spelling of the heading.
%% \begin{acks}

%% \end{acks}

\bibliographystyle{ACM-Reference-Format}
\bibliography{references}

%% \section{Research Methods}
%% \subsection{Part One}
\section*{Appendix}
%\subsection*{Detailed Cross-LLM Results for Baseline 0}

\begin{table*}[t]
\centering
\small
\caption{Baseline 0 vs. CHARGE using GPT-4.1.}
\label{tab:Baseline0_GPT}
\begin{tabular}{lcccccccc}
\hline
Module & CWE & Bug &
\multicolumn{2}{c}{Detection Result} &
\multicolumn{2}{c}{Compilable} & 
\multicolumn{2}{c}{Unreachable}  \\
\cline{4-9}
&&&
Baseline0 & CHARGE &
Baseline0 & CHARGE & 
Baseline0 & CHARGE \\
\hline

sha256\_wrapper & 1239 & P36 & \xmark & \cmark & 3/3 & 7/7 & 0/3 & 0/7\\
aes0\_wrapper & 226 & P39 & \xmark & \cmark & 3/3 & 44/44 & 0/3 & 0/44 \\
aes0\_wrapper & 1258 & P46 & \xmark & \cmark &
\multirow{2}{*}{5/5} &
\multirow{2}{*}{25/25} &
\multirow{2}{*}{0/5} & 
\multirow{2}{*}{0/25} \\
& & P47 & \xmark & \cmark & & \\
dmi\_jtag & 1245 & P2 & \xmark & \cmark & 
\multirow{2}{*}{4/5} &
\multirow{2}{*}{33/33} &
\multirow{2}{*}{0/4} & 
\multirow{2}{*}{0/33} \\
& & P84 & \xmark & \xmark & & \\
csr\_regfile & 1262 & P18 & \xmark & \xmark & 3/3 & 21/34 & 0/3 & 8/21 \\
reglk\_wrapper & 1234 & P48 & \xmark & \cmark & 3/3 & 14/14 & 0/3 & 0/14 \\
reglk\_wrapper & 1232 & P35 & \xmark & \cmark & 5/5 & 19/20 & 0/5 & 5/19\\
rsa\_wrapper & 226 & P95 & \cmark & \cmark & 6/6 & 28/28 & 0/6 & 4/28\\
acct\_wrapper & 276 & P42 & \xmark & \cmark & 0/3 & 6/9 & N/A & 1/6\\
dma & 1245 & P57 & \xmark & \cmark & 3/3 & 33/34 & 0/3 & 3/33\\
riscv\_peripherals & 1310 & P96 & \xmark & \xmark & 3/3 & 13/21 & 0/3 & 1/13\\
aes\_192 & 1240 & P14 & \xmark & \cmark & 3/3 & 39/48 & 0/3 & 0/39\\

\hline
Results & & & 1/14 & 11/14 & 41/45 & 282/317 & 0/41 & 22/282 \\
\hline
\end{tabular}
\end{table*}

\begin{table*}[h]
\centering
\small
\caption{Baseline 0 vs. CHARGE using Gemini 2.5 Pro}
\label{tab:Baseline0_Gemini}
\begin{tabular}{lcccccccc}
\hline
Module & CWE & Bug &
\multicolumn{2}{c}{Detection Result} &
\multicolumn{2}{c}{Compilable} & 
\multicolumn{2}{c}{Unreachable}  \\
\cline{4-9}
&&&
Baseline0 & CHARGE &
Baseline0 & CHARGE & 
Baseline0 & CHARGE \\
\hline

sha256\_wrapper & 1239 & P36 & \xmark & \cmark & 8/8 & 24/24 & 0/8 & 0/24\\
aes0\_wrapper & 226 & P39 & \xmark & \cmark & 6/6 & 65/65 & 0/6 & 0/65 \\
aes0\_wrapper & 1258 & P46 & \cmark & \cmark &
\multirow{2}{*}{6/6} &
\multirow{2}{*}{63/63} &
\multirow{2}{*}{0/6} & 
\multirow{2}{*}{1/63} \\
& & P47 & \cmark & \cmark & & \\
dmi\_jtag & 1245 & P2 & \xmark & \cmark & 
\multirow{2}{*}{5/9} &
\multirow{2}{*}{65/73} &
\multirow{2}{*}{2/5} & 
\multirow{2}{*}{7/65} \\
& & P84 & \xmark & \xmark & & \\
csr\_regfile & 1262 & P18 & \xmark & \xmark & 27/28 & 60/60 & 6/27 & 2/60 \\
reglk\_wrapper & 1234 & P48 & \cmark & \cmark & 3/3 & 14/14 & 0/3 & 1/14 \\
reglk\_wrapper & 1232 & P35 & \cmark & \cmark & 3/3 & 7/7 & 0/3 & 0/7\\
rsa\_wrapper & 226 & P95 & \xmark & \cmark & 3/3 & 79/79 & 0/3 & 4/79\\
acct\_wrapper & 276 & P42 & \cmark & \cmark & 3/3 & 4/7 & 0/3 & 0/4\\
dma & 1245 & P57 & \cmark & \cmark & 24/24 & 69/69 & 1/24 & 10/69\\
riscv\_peripherals & 1310 & P96 & \xmark & \cmark & 0/4 & 12/16 & N/A & 4/12\\
aes\_192 & 1240 & P14 & \xmark & \xmark & 9/9 & 27/27 & 0/9 & 0/27\\

\hline
Results & & & 6/14 & 11/14 & 97/106 & 489/504 & 9/97 & 29/489 \\
\hline
\end{tabular}
\end{table*}

\begin{table*}[h]
\centering
\small
\caption{Baseline 0 vs. CHARGE using Claude Sonnet 4.6}
\label{tab:Baseline0_Claude}
\begin{tabular}{lcccccccc}
\hline
Module & CWE & Bug &
\multicolumn{2}{c}{Detection Result} &
\multicolumn{2}{c}{Compilable} & 
\multicolumn{2}{c}{Unreachable}  \\
\cline{4-9}
&&&
Baseline0 & CHARGE &
Baseline0 & CHARGE & 
Baseline0 & CHARGE \\
\hline

sha256\_wrapper & 1239 & P36 & \xmark & \xmark & 30/30 & 74/74 & 3/30 & 19/74\\
aes0\_wrapper & 226 & P39 & \xmark & \cmark & 24/31 & 108/108 & 0/24 & 13/108 \\
aes0\_wrapper & 1258 & P46 & \cmark & \cmark &
\multirow{2}{*}{21/21} &
\multirow{2}{*}{82/82} &
\multirow{2}{*}{0/21} & 
\multirow{2}{*}{0/82} \\
& & P47 & \cmark & \cmark & & \\
dmi\_jtag & 1245 & P2 & \xmark & \cmark & 
\multirow{2}{*}{51/52} &
\multirow{2}{*}{131/147} &
\multirow{2}{*}{5/51} & 
\multirow{2}{*}{10/131} \\
& & P84 & \xmark & \xmark & & \\
csr\_regfile & 1262 & P18 & \xmark & \cmark & 66/68 & 118/134 & 0/66 & 6/118 \\
reglk\_wrapper & 1234 & P48 & \xmark & \cmark & 16/16 & 21/22 & 0/16 & 0/21 \\
reglk\_wrapper & 1232 & P35 & \xmark & \cmark & 16/16 & 18/18 & 0/16 & 0/18\\
rsa\_wrapper & 226 & P95 & \xmark & \cmark & 37/37 & 125/125 & 0/37 & 24/125\\
acct\_wrapper & 276 & P42 & \xmark & \cmark & 10/11 & 55/55 & 0/10 & 18/55\\
dma & 1245 & P57 & \xmark & \cmark & 47/47 & 133/133 & 2/47 & 19/133\\
riscv\_peripherals & 1310 & P96 & \cmark & \cmark & 12/12 & 98/98 & 0/12 & 8/98\\
aes\_192 & 1240 & P14 & \xmark & \xmark & 39/39 & 295/295 & 0/39 & 0/295\\

\hline
Results & & & 3/14 & 11/14 & 369/380 & 1258/1291 & 10/369 & 117/1258 \\
\hline
\end{tabular}
\end{table*}

%%%%%%%%%%%%%%%%%%%%%%%%%%%%%%%%%%%%%%%%%%%%%%%%%%%%%%%
%\subsection*{Detailed Cross-LLM Results for Baseline 1}

\begin{table*}[t]
\centering
\small
\caption{Asset identification results using GPT-4.1.}
\label{tab:Baseline1_GPT}
\begin{tabular}{lcccccc}
\hline
Module & CWE & Bug & Core Assets & Hierarchy+3Tuple & 3Tuple & None \\
\hline
sha256\_wrapper      & 1239 & P36 & 2 & 2   & 2   & 2   \\
aes0\_wrapper        & 226  & P39 & 2 & 1.5 & 1.5 & 2   \\
aes0\_wrapper        & 1258 & P46 & 4 & 3   & 1   & 0.5 \\
aes0\_wrapper        & 1258 & P47 & 4 & 3   & 1   & 0.5 \\
dmi\_jtag            & 1245 & P2  & 3 & 2.5 & 2   & 3   \\
dmi\_jtag            & 1245 & P84 & 2 & 1   & 1   & 1   \\
csr\_regfile         & 1262 & P18 & 6 & 1.5 & 1.5 & 1.5 \\
reglk\_wrapper       & 1234 & P48 & 3 & 3   & 1   & 1.5 \\
reglk\_wrapper       & 1232 & P35 & 4 & 4   & 4   & 3   \\
rsa\_wrapper         & 226  & P95 & 3 & 3   & 3   & 1   \\
acct\_wrapper        & 276  & P42 & 3 & 3   & 2.5 & 1   \\
dma                  & 1245 & P57 & 2 & 1.5 & 1   & 1   \\
riscv\_peripherals   & 1310 & P96 & 2 & 0   & 0   & 0.5 \\
aes\_192             & 1240 & P14 & 1 & 1   & 0   & 0   \\
\hline
Total Score & & & 41 & 30 & 21.5 & 18.5 \\
Accuracy & & & & 73.2\% & 52.4\% & 45.1\% \\
\hline
\end{tabular}
\end{table*}

\begin{table*}[h]
\centering
\small
\caption{Asset identification results using Gemini 2.5 Pro.}
\begin{tabular}{lcccccc}
\hline
Module & CWE & Bug & Core Assets & Hierarchy+3Tuple & 3Tuple & None \\
\hline
sha256\_wrapper      & 1239 & P36 & 2 & 2 & 1.5 & 1 \\
aes0\_wrapper        & 226  & P39 & 2 & 2 & 1.5 & 0 \\
aes0\_wrapper        & 1258 & P46 & 4 & 4 & 2 & 2.5 \\
aes0\_wrapper        & 1258 & P47 & 4 & 4 & 2 & 2.5 \\
dmi\_jtag            & 1245 & P2  & 3 & 3 & 2 & 3 \\
dmi\_jtag            & 1245 & P84 & 2 & 2 & 1.5 & 1 \\
csr\_regfile         & 1262 & P18 & 6 & 3 & 2.5 & 0 \\
reglk\_wrapper       & 1234 & P48 & 3 & 2.5 & 2 & 2.5 \\
reglk\_wrapper       & 1232 & P35 & 4 & 4 & 4 & 4 \\
rsa\_wrapper         & 226  & P95 & 3 & 3 & 2.5 & 1 \\
acct\_wrapper        & 276  & P42 & 3 & 3 & 3 & 3 \\
dma                  & 1245 & P57 & 2 & 2 & 1.5 & 2 \\
riscv\_peripherals   & 1310 & P96 & 2 & 2 & 1 & 2 \\
aes\_192             & 1240 & P14 & 1 & 1 & 1 & 0 \\
\hline
Total Score& & & 41 & 37.5 & 28 & 24.5 \\
Accuracy & & & & 91.5\% & 68.3\% & 59.8\% \\
\hline
\end{tabular}
\end{table*}

\begin{table*}[h]
\centering
\small
\caption{Asset identification results using Claude Sonnet 4.6.}
\begin{tabular}{lcccccc}
\hline
Module & CWE & Bug & Core Assets & Hierarchy+3Tuple & 3Tuple & None \\
\hline
sha256\_wrapper      & 1239 & P36 & 2 & 2 & 2 & 2 \\
aes0\_wrapper        & 226  & P39 & 2 & 2 & 1 & 1 \\
aes0\_wrapper        & 1258 & P46 & 4 & 4 & 4 & 4 \\
aes0\_wrapper        & 1258 & P47 & 4 & 4 & 4 & 4 \\
dmi\_jtag            & 1245 & P2  & 3 & 3 & 3 & 3 \\
dmi\_jtag            & 1245 & P84 & 2 & 2 & 1 & 1.5 \\
csr\_regfile         & 1262 & P18 & 6 & 4 & 5 & 3.5 \\
reglk\_wrapper       & 1234 & P48 & 3 & 3 & 2 & 2 \\
reglk\_wrapper       & 1232 & P35 & 4 & 4 & 3.5 & 3 \\
rsa\_wrapper         & 226  & P95 & 3 & 3 & 2.5 & 1 \\
acct\_wrapper        & 276  & P42 & 3 & 3 & 3 & 3 \\
dma                  & 1245 & P57 & 2 & 2 & 1.5 & 1 \\
riscv\_peripherals   & 1310 & P96 & 2 & 1 & 2 & 2 \\
aes\_192             & 1240 & P14 & 1 & 1 & 1 & 1 \\
\hline
Total Score & & & 41 & 38 & 35.5 & 32 \\
Accuracy & & & & 92.7\% & 86.6\% & 78.0\% \\
\hline
\end{tabular}
\end{table*}

\end{document}